\documentclass[11pt,preprint,aps,showkeys]{revtex4}
\usepackage{amsmath}
\usepackage{amsfonts}
\usepackage{amssymb}
\usepackage{natbib}
\usepackage{graphicx}
\usepackage{epstopdf}
\usepackage{subfigure}
\usepackage{mathrsfs}
\usepackage{dcolumn}
\usepackage{bm}
\usepackage[section]{placeins}
\usepackage{xcolor}
\usepackage{hyperref}

\begin{document}

\title{Second-order multipole response of two nearby incoherent sources in wave-optical imaging of a Schwarzschild black hole}

\author{Wenfu Cao}\email{202411100001@stu.ujn.edu.cn}

\author{Hongsheng Zhang}\email{corresponding author: sps\_zhanghs$@$ujn.edu.cn}

\affiliation{School of Physics and Technology, University of Jinan, 336 West Road of Nan Xinzhuang, Jinan, Shandong 250022, China}
\author{Zong-Kuan Guo}
\email{guozk@itp.ac.cn }
\affiliation{Institute of Theoretical Physics, Chinese Academy of Sciences, Beijing 100190, China}
 \author{Rong-Gen Cai}
\email{cairg@itp.ac.cn }
\affiliation{Zhejiang Key Laboratory of Extreme Universe, Ningbo University, Ningbo 315211, China}
\affiliation{Institute of Fundamental Physics and Quantum Technology, Ningbo University, Ningbo 315211, China}

\begin{abstract}
We study wave-optical imaging of two nearby, mutually incoherent point sources by a Schwarzschild black hole. Using the spherical-harmonic addition theorem, we construct partial-wave fields for arbitrary source directions and form images through a finite-aperture Fourier transform. To isolate the binary structure, we compare the binary image with that of a single source of equal total brightness placed at the brightness centroid. Expanding about the centroid removes the first-order term exactly, so the leading structural correction is controlled by the second central moment, $\mu_2=4a^2q/(1+q)^2$. Numerical calculations confirm the expected dependence of the residual on source half-separation and brightness ratio. We then study its wavelength dependence. At long wavelengths, the imaging kernel varies little across the source separation, making the binary nearly indistinguishable from the centroid reference. As the wavelength decreases, finer fringes enhance the residual, which grows by about a factor of three over $3\leq M\omega\leq12$ ($2.09\geq\lambda/M\geq0.52$), even though the pair remains unresolved in the Rayleigh sense. Meanwhile, the second-order approximation gradually breaks down: its relative error first reaches unity for $17<M\omega<18$ ($0.349<\lambda/M<0.370$), indicating the need for higher-order source moments. These results demonstrate that the distinguishability of a nearby binary is intrinsically chromatic and identify the wavelength range in which the binary is both detectable through its residual and accurately described at second order.
\end{abstract}

\keywords{black-hole wave optics; Schwarzschild black hole; incoherent binary sources; partial-wave method; multipole expansion}

\maketitle

\section{Introduction}

Wave propagation near a black hole is affected simultaneously by strong gravitational deflection, absorption, diffraction, and interference.
Most studies of black-hole imaging employ the geometric-optics approximation and describe radiation by null geodesics \cite{EventHorizonTelescope:2019dse,EventHorizonTelescope:2022wkp,Wei:2013kza,Guo:2020zmf,Zeng:2020dco,Liu:2020ola,Wang:2017hjl,Huang:2018rfn,Huang:2024gtu,Huang:2025xqd,Gralla:2019xty,Hu:2022lek,Hu:2023bzy,Hu:2023pyd}.
This approximation efficiently captures the shadow boundary and critical curves when the wavelength is much shorter than the characteristic scale of the black hole. When the two scales become comparable, however, diffraction and interference can no longer be neglected. One must instead solve a wave equation in the curved spacetime and reconstruct the corresponding image by wave-optical methods. Wave effects in gravitational lensing have long been known to produce frequency-dependent amplification and interference signatures that are absent in geometric optics \cite{Deguchi:1986zz,Takahashi:2003ix}.

Wave-optical imaging of black holes has been investigated in a series of works. Earlier partial-wave studies established the characteristic absorption, forward scattering, backward glory, and oscillatory interference of waves scattered by nonrotating black holes \cite{Sanchez:1976fcl,Sanchez:1977vz,Crispino:2009ki}. Kanai et al. studied the spatial distribution of scalar waves emitted by a point source in a Schwarzschild spacetime and reconstructed black-hole images by a finite-aperture Fourier transform \cite{Kanai:2013rga}. Nambu et al. subsequently analyzed wave scattering in the Schwarzschild geometry, including the interference between direct and orbiting waves \cite{Nambu:2015aea}. The partial-wave method was later extended to general static and spherically symmetric compact objects, allowing wave-optical images of black holes and horizonless objects to be compared \cite{Nambu:2019sqn}. These studies demonstrate that even a single point source can produce bright rings, central spots, multiple peaks, and intricate diffraction fringes on the image plane.

Wave-optical imaging has also been developed in asymptotically anti-de Sitter spacetimes. Hashimoto et al. proposed reconstructing holographic Einstein rings from boundary response functions \cite{Hashimoto:2019jmw,Hashimoto:2018okj}. Related holographic imaging studies have considered charged and higher-curvature AdS black holes \cite{Liu:2022cev,Zeng:2023zlf}. More recently, an exact treatment based on Heun functions was applied to wave-optical imaging of a Kerr--de Sitter black hole \cite{Willenborg:2023ixu}, while spin-dependent wave-optical effects have been investigated for gravitational waves lensed by a Kerr black hole \cite{Kubota:2024zkv}. These developments have broadened the scope of black-hole wave optics. The image structures produced by two nearby incoherent point sources, however, have not received the same systematic attention.

A natural approach to a two-source problem is to count bright peaks or to use conventional criteria such as the Rayleigh or Sparrow criteria \cite{Rayleigh1879,Sparrow1916}. Such criteria are not directly reliable for identifying the number of physical sources in black-hole wave optics. The black hole is itself a complicated scattering system, and a single point source may already generate several image peaks, rings, and interference sidelobes. Consequently, two peaks in an image do not necessarily imply two physical sources, while the absence of a visually separated pair of peaks does not imply that all information about the binary source has been lost.

A related lesson has recently emerged in strong-lensing searches for
gravitational waves.  Ali et al. showed that comparing lensed
signals with multi-image waveform templates can substantially reduce the
waveform mismatch relative to an insufficient lower-image-number model
\cite{Ali:2025cqo}.  Although their observable and lens model differ
from those considered here, their result illustrates the usefulness of
model-based residuals for identifying unresolved multiple-image structure.

This observation suggests comparing the binary-source image with an effective single-source image having the same total brightness and the same overall position. Such a comparison removes changes associated only with the total flux and the displacement of the image, thereby isolating image information that originates from the internal spatial distribution of the source.

In this work, we investigate the wave-optical imaging of two nearby,
mutually incoherent point sources by a Schwarzschild black hole. We place
a reference single source with the same total source brightness at the
brightness centroid of the pair and adopt it as the zeroth-order model.
Expanding about this centroid eliminates the first-central-moment
contribution exactly, while the second central moment gives the leading
structural correction to the reference image. We refer to the
corresponding image-plane contribution as the second-order multipole
response. Using the full partial-wave solution and Fourier imaging, we
first test the dependence on half-separation and brightness ratio
predicted by the second-order expansion. We then fix the source
parameters and investigate how frequency affects the morphology of the
second-order response, the strength of the full image residual, and the
relative error of the second-order approximation, thereby quantifying its
range of validity for the parameters considered.

The remainder of this paper is organized as follows.
Section~\ref{sec:two} establishes the two-point-source partial-wave model
in a Schwarzschild spacetime. Section~\ref{sec:three} introduces the
construction of the lens-plane field and the wave-optical imaging method.
Section~\ref{sec:four} derives the brightness-centroid expansion, the
second-order multipole response, and the associated parameter scalings,
and discusses the conditions for the expansion to apply.
Section~\ref{sec:numerical_verification} presents the numerical tests:
Sec.~\ref{subsec:second_order_response} examines the second-order response
and its dependence on half-separation and brightness ratio, while
Sec.~\ref{subsec:frequency_dependence} investigates frequency dependence
and the validity of the second-order approximation.
Section~\ref{sec:conclusions} summarizes the main results and discusses
possible extensions.

\section{Two-point-source partial-wave model}\label{sec:two}

The purpose of this section is to construct the frequency-domain field produced by two off-axis point sources in a Schwarzschild spacetime. We first specify the source geometry and the sourced scalar-wave equation. We then use the spherical-harmonic addition theorem to separate the common radial propagation from the source-dependent angular factors. Because the two sources have the same radial coordinate, only one radial Green function is required for each multipole number $l$. The resulting expression will provide the lens-plane fields that are imaged separately in Sec.~III.

\subsection{Sourced scalar field and two-source geometry}

We consider a massless scalar field propagating in a Schwarzschild spacetime,
\begin{equation}
 ds^{2}=-f(r)dt^{2}+\frac{dr^{2}}{f(r)}
 +r^{2}\left(d\theta^{2}+\sin^{2}\theta\,d\phi^{2}\right),
 \qquad
 f(r)=1-\frac{2M}{r}.
 \label{eq:schwarzschild_metric}
\end{equation}
For a point source at $\mathbf{x}_{s}$, the sourced Klein--Gordon equation is
\begin{equation}
 \Box\Psi_{s}=-\mathcal{S}_{s}(t)
 \delta^{(3)}(\mathbf{x};\mathbf{x}_{s}),
 \label{eq:sourced_kg}
\end{equation}
where the invariant spatial delta distribution is normalized by
\begin{equation}
 \int dr\,d\theta\,d\phi\,\sqrt{-g}\,
 \delta^{(3)}(\mathbf{x};\mathbf{x}_{s})=1.
 \label{eq:delta_normalization}
\end{equation}
Since $\sqrt{-g}=r^{2}\sin\theta$ for the Schwarzschild metric, it can be written in the coordinate form
\begin{equation}
 \delta^{(3)}(\mathbf{x};\mathbf{x}_{s})
 =\frac{\delta(r-r_{s})\delta(\theta-\theta_{s})
 \delta(\phi-\phi_{s})}{r^{2}\sin\theta}.
 \label{eq:invariant_delta}
\end{equation}
For a monochromatic source, we set
\begin{equation}
 \Psi_{s}(t,r,\Omega)=e^{-i\omega t}\Phi_{s}(r,\Omega),
 \qquad \Omega=(\theta,\phi).
 \label{eq:monochromatic_field}
\end{equation}

The two sources considered in this work are located at
\begin{equation}
 \mathbf{r}_{s1}=(a,0,-z_{0}),
 \qquad
 \mathbf{r}_{s2}=(-a,0,-z_{0}).
 \label{eq:source_positions}
\end{equation}
They have the same radial coordinate
\begin{equation}
 r_{s}=\sqrt{z_{0}^{2}+a^{2}},
 \label{eq:source_radius}
\end{equation}
but different angular directions,
\begin{equation}
 \mathbf{n}_{s1}=\frac{1}{r_{s}}(a,0,-z_{0}),
 \qquad
 \mathbf{n}_{s2}=\frac{1}{r_{s}}(-a,0,-z_{0}).
 \label{eq:source_directions}
\end{equation}
Consequently, the two sources share the same radial propagation functions, while their angular dependences are different.

\subsection{Partial-wave Green-function solution}

The angular delta distribution for a source in an arbitrary direction $\Omega_s$ admits the spherical-harmonic expansion
\begin{equation}
 \delta^{(2)}(\Omega-\Omega_{s})
 =\sum_{l=0}^{\infty}\sum_{m=-l}^{l}
 Y_{lm}(\Omega)Y_{lm}^{*}(\Omega_{s}).
 \label{eq:angular_delta_expansion}
\end{equation}
Because the Schwarzschild background is spherically symmetric, the radial operator depends on $l$ but not on $m$. The sum over $m$ can therefore be performed by using the spherical-harmonic addition theorem,
\begin{equation}
 \sum_{m=-l}^{l}Y_{lm}(\Omega)Y_{lm}^{*}(\Omega_{s})
 =\frac{2l+1}{4\pi}P_{l}(\mathbf{n}\cdot\mathbf{n}_{s}),
 \label{eq:addition_theorem}
\end{equation}
where
\begin{equation}
 \mathbf{n}=(\sin\theta\cos\phi,
 \sin\theta\sin\phi,\cos\theta)
 \label{eq:field_direction}
\end{equation}
is the unit vector pointing from the black hole to the field point. The frequency-domain field generated by source $s$ can then be expressed as
\begin{equation}
 \Phi_{s}(r,\Omega)
 =\frac{\mathcal{J}_{s}}{r}
 \sum_{l=0}^{\infty}\frac{2l+1}{4\pi}
 G_{l}(r,r_{s})P_{l}(\mathbf{n}\cdot\mathbf{n}_{s}),
 \label{eq:single_source_partial_wave}
\end{equation}
where $\mathcal{J}_{s}$ contains the Fourier amplitude and normalization of the source. Equation~\eqref{eq:single_source_partial_wave} does not discard the azimuthal modes; rather, all $m$ modes have been summed exactly through Eq.~\eqref{eq:addition_theorem}.

For the source geometry in Eq.~\eqref{eq:source_positions}, the two individual fields are
\begin{align}
 \Phi_{1}(r,\Omega)
 &=\frac{\mathcal{J}_{1}}{r}
 \sum_{l=0}^{\infty}\frac{2l+1}{4\pi}
 G_{l}(r,r_{s})P_{l}(\mathbf{n}\cdot\mathbf{n}_{s1}),
 \label{eq:source1_field}\\
 \Phi_{2}(r,\Omega)
 &=\frac{\mathcal{J}_{2}}{r}
 \sum_{l=0}^{\infty}\frac{2l+1}{4\pi}
 G_{l}(r,r_{s})P_{l}(\mathbf{n}\cdot\mathbf{n}_{s2}).
 \label{eq:source2_field}
\end{align}

To complete the partial-wave solution in Eq.~\eqref{eq:single_source_partial_wave}, it remains to determine the radial response $G_l(r,r_s)$. Introducing the tortoise coordinate
\begin{equation}
 \frac{dr_{*}}{dr}=\frac{1}{f(r)},
 \qquad
 r_{*}=r+2M\ln\left(\frac{r}{2M}-1\right),
 \label{eq:tortoise_coordinate}
\end{equation}
the radial Green function satisfies
\begin{equation}
 \left[\frac{d^{2}}{dr_{*}^{2}}+\omega^{2}-V_{l}(r)\right]
 G_{l}(r,r_{s})=\delta(r_{*}-r_{*s}),
 \label{eq:radial_green_equation}
\end{equation}
with the Schwarzschild effective potential
\begin{equation}
 V_{l}(r)=f(r)\left[\frac{l(l+1)}{r^{2}}+\frac{2M}{r^{3}}\right].
 \label{eq:effective_potential}
\end{equation}
Let $u_{l}^{\mathrm{in}}$ denote the homogeneous solution that is purely ingoing at the event horizon,
\begin{equation}
 u_{l}^{\mathrm{in}}\sim e^{-i\omega r_{*}},
 \qquad r_{*}\rightarrow-\infty,
 \label{eq:horizon_boundary}
\end{equation}
and let $u_{l}^{\mathrm{up}}$ denote the solution that is purely outgoing at spatial infinity,
\begin{equation}
 u_{l}^{\mathrm{up}}\sim e^{+i\omega r_{*}},
 \qquad r_{*}\rightarrow+\infty.
 \label{eq:infinity_boundary}
\end{equation}
For the numerical integration, these boundary conditions are implemented
by the standard local series forms
\begin{align}
 u_l^{\mathrm{in}}(r)
 &= (r-2M)^{-2iM\omega}
    \sum_{n=0}^{N_H}a_n(r-2M)^n,
    \qquad r\rightarrow2M,
    \label{eq:horizon_frobenius_series}\\
 u_l^{\mathrm{up}}(r)
 &= e^{i\omega r}r^{2iM\omega}
    \sum_{n=0}^{N_\infty}\frac{b_n}{r^n},
    \qquad r\rightarrow\infty.
    \label{eq:infinity_asymptotic_series}
\end{align}
The first expression is a Frobenius expansion about the regular singular
point at the event horizon, while the second is an asymptotic expansion
about the irregular singular point at infinity.  The coefficients $a_n$
and $b_n$ follow recursively after substitution into the homogeneous
radial equation; $a_0$ and $b_0$ set arbitrary normalizations.  Truncating
the two series and differentiating them provides the pairs $(u_l,du_l/dr)$
used to start the inward and outward integrations.  This is the standard
series construction used in black-hole quasinormal-mode calculations
\cite{Leaver1985}.

The radial Green function is then
\begin{equation}
 G_{l}(r,r_{s})
 =\frac{u_{l}^{\mathrm{in}}(r_{<})u_{l}^{\mathrm{up}}(r_{>})}{W_{l}},
 \qquad
 r_{<}=\min(r,r_{s}),\quad r_{>}=\max(r,r_{s}),
 \label{eq:radial_green_solution}
\end{equation}
where
\begin{equation}
 W_{l}=u_{l}^{\mathrm{in}}\frac{du_{l}^{\mathrm{up}}}{dr_{*}}
 -u_{l}^{\mathrm{up}}\frac{du_{l}^{\mathrm{in}}}{dr_{*}}
 \label{eq:wronskian}
\end{equation}
is the Wronskian. This construction guarantees continuity of $G_l$ at the source radius and the unit derivative jump required by the delta distribution in Eq.~\eqref{eq:radial_green_equation}.

\subsection{Incoherent binary and single-source limit}

The two sources are assumed to be statistically independent and mutually incoherent. Their complex amplitudes therefore satisfy
\begin{equation}
 \left\langle\mathcal{J}_{1}\mathcal{J}_{2}^{*}\right\rangle=0.
 \label{eq:incoherent_condition}
\end{equation}
Accordingly, $\Phi_1$ and $\Phi_2$ are propagated and imaged separately. Their image-plane intensities, rather than their complex amplitudes, are added. The imaging operation and the resulting incoherent sum will be introduced in Sec.~III.

As a consistency check, consider the limit $a\rightarrow0$. Both source directions then approach the negative $z$ axis,
\begin{equation}
 \mathbf{n}_{s1},\mathbf{n}_{s2}\rightarrow(0,0,-1),
 \label{eq:axis_source_limit}
\end{equation}
so that
\begin{equation}
 \mathbf{n}\cdot\mathbf{n}_{s}=-\cos\theta,
 \qquad
 P_{l}(-\cos\theta)=(-1)^{l}P_{l}(\cos\theta).
 \label{eq:legendre_axis_limit}
\end{equation}
The off-axis expression in Eq.~\eqref{eq:single_source_partial_wave} therefore reduces to the familiar axisymmetric single-source expansion. This limit provides a direct validation of the two-source angular construction.

\section{Lens-plane field and wave-optical imaging}\label{sec:three}

Having obtained the partial-wave field of each source, we now convert it into an observable wave-optical image. The purpose of this section is to map the scattered field from the observation sphere to a finite circular lens and then to the image plane. We first construct the lens-plane field for an arbitrary observation direction $(\theta_0,\phi_0)$. We then apply the Fourier imaging integral to each source separately and add the resulting intensities according to the incoherent-source condition.

\subsection{Lens geometry and incident complex field}

We place the center of the lens at the observation direction
$(\theta_{0},\phi_{0})$. The corresponding radial unit vector is
\begin{equation}
 \mathbf{n}_{0}=
 (\sin\theta_{0}\cos\phi_{0},
  \sin\theta_{0}\sin\phi_{0},
  \cos\theta_{0}).
 \label{eq:lens_center_direction}
\end{equation}
Two orthonormal vectors tangent to the observation sphere at
$\mathbf{n}_{0}$ are chosen as
\begin{align}
 \mathbf{e}_{X}&=
 (\cos\theta_{0}\cos\phi_{0},
  \cos\theta_{0}\sin\phi_{0},
  -\sin\theta_{0}),
 \label{eq:lens_basis_x}\\
 \mathbf{e}_{Y}&=
 (-\sin\phi_{0},\cos\phi_{0},0).
 \label{eq:lens_basis_y}
\end{align}
They obey
$\mathbf{n}_{0}\cdot\mathbf{e}_{X}
=\mathbf{n}_{0}\cdot\mathbf{e}_{Y}
=\mathbf{e}_{X}\cdot\mathbf{e}_{Y}=0$.
The opposite sign convention for either transverse basis vector merely
reverses the corresponding image axis and has no physical consequence.

Let $(X,Y)$ denote local transverse coordinates on the lens. A point on
the lens is associated with the unit direction
\begin{equation}
 \mathbf{n}_{\mathrm{lens}}(X,Y)
 =\sqrt{1-\frac{X^{2}+Y^{2}}{r_{\mathrm{obs}}^{2}}}\,
  \mathbf{n}_{0}
 +\frac{X}{r_{\mathrm{obs}}}\mathbf{e}_{X}
 +\frac{Y}{r_{\mathrm{obs}}}\mathbf{e}_{Y},
 \label{eq:lens_direction}
\end{equation}
where $r_{\mathrm{obs}}$ is the observation radius. Equation~\eqref{eq:lens_direction}
ensures $|\mathbf{n}_{\mathrm{lens}}|=1$. The circular aperture is restricted to
$X^{2}+Y^{2}\leq d^{2}$, where $d<r_{\mathrm{obs}}$ is the lens radius.
Thus $(\theta_{0},\phi_{0})$ fixes the center of the field of view, while
$(X,Y)$ samples the neighboring propagation directions collected by the lens.

With this local geometry, the angular separation between a lens point and
source $s$ is determined directly by
\begin{equation}
 \cos\gamma_{s}(X,Y)
 =\mathbf{n}_{\mathrm{lens}}(X,Y)\cdot\mathbf{n}_{s}.
 \label{eq:lens_source_angle}
\end{equation}
Evaluating Eq.~\eqref{eq:single_source_partial_wave} at
$r=r_{\mathrm{obs}}$ gives the complex field incident on the lens,
\begin{equation}
 \Phi_{\mathrm{lens},s}(X,Y)
 =\frac{\mathcal{J}_{s}}{r_{\mathrm{obs}}}
 \sum_{l=0}^{\infty}\frac{2l+1}{4\pi}
 G_{l}(r_{\mathrm{obs}},r_{s})
 P_{l}\!\left[\cos\gamma_{s}(X,Y)\right]
 \mathcal{A}_{d}(X,Y),
 \label{eq:lens_field}
\end{equation}
where the aperture function is
\begin{equation}
 \mathcal{A}_{d}(X,Y)=
 \begin{cases}
  1, & X^{2}+Y^{2}\leq d^{2},\\
  0, & X^{2}+Y^{2}>d^{2}.
 \end{cases}
 \label{eq:aperture_function}
\end{equation}
The construction retains both the amplitude and phase of the scattered
wave. In particular, the two sources are evaluated with their own
$\mathbf{n}_{s}$ and therefore generate two distinct complex fields on
the same lens.

\subsection{Fourier imaging of the incoherent binary}

In the thin-lens and paraxial approximations, the complex image amplitude
generated by source $s$ is proportional to the Fourier transform of the
lens-plane field,
\begin{equation}
 \Phi_{\mathrm{I},s}(X_{\mathrm{I}},Y_{\mathrm{I}})
 \propto
 \int_{X^{2}+Y^{2}\leq d^{2}}dX\,dY\,
 \Phi_{\mathrm{lens},s}(X,Y)
 \exp\!\left[-\frac{i\omega}{f}
 (XX_{\mathrm{I}}+YY_{\mathrm{I}})\right],
 \label{eq:fourier_imaging_physical}
\end{equation}
where $f$ is the focal length and $(X_{\mathrm{I}},Y_{\mathrm{I}})$
are Cartesian coordinates on the image plane. Introducing the image
angles
\begin{equation}
 \alpha=\frac{X_{\mathrm{I}}}{f},
 \qquad
 \beta=\frac{Y_{\mathrm{I}}}{f},
 \label{eq:image_angles}
\end{equation}
we obtain
\begin{equation}
 \Phi_{\mathrm{I},s}(\alpha,\beta)
 \propto
 \int_{X^{2}+Y^{2}\leq d^{2}}dX\,dY\,
 \Phi_{\mathrm{lens},s}(X,Y)
 e^{-i\omega(\alpha X+\beta Y)}.
 \label{eq:fourier_imaging_angular}
\end{equation}
In the numerical implementation this integral is evaluated by a
two-dimensional fast Fourier transform. If $(f_{x},f_{y})$ denote the
spatial frequencies returned by the transform, the corresponding image
angles are
\begin{equation}
 \alpha=\frac{2\pi f_{x}}{\omega},
 \qquad
 \beta=\frac{2\pi f_{y}}{\omega}.
 \label{eq:fft_image_coordinates}
\end{equation}

The Fourier operation must be applied to the two source fields separately.
For two coherent sources, the image intensity would contain the
interference term between $\Phi_{\mathrm{I},1}$ and
$\Phi_{\mathrm{I},2}$. In the present problem the sources are mutually
incoherent. Using Eq.~\eqref{eq:incoherent_condition}, the time-averaged
cross term vanishes and the binary-source image becomes
\begin{align}
 I_{\mathrm{binary}}(\alpha,\beta)
 &=\left\langle
 \left|\Phi_{\mathrm{I},1}+\Phi_{\mathrm{I},2}\right|^{2}
 \right\rangle \nonumber\\
 &=\left|\Phi_{\mathrm{I},1}(\alpha,\beta)\right|^{2}
  +\left|\Phi_{\mathrm{I},2}(\alpha,\beta)\right|^{2}.
 \label{eq:incoherent_image_sum}
\end{align}
The source brightness is proportional to
$F_{s}=|\mathcal{J}_{s}|^{2}$. Consequently, the two complex lens fields
must be transformed separately before their image-plane intensities are
added; adding the lens-plane amplitudes first would instead describe a
coherent source pair.

All quantitative comparisons below are performed with the unnormalized
intensities in Eq.~\eqref{eq:incoherent_image_sum}, so that the physical
brightness ratio is preserved. A normalization by the maximum intensity
is applied only when displaying an image and is not used in the
brightness-centroid expansion or in the image-residual measure.

\section{Brightness-centroid expansion and second-order multipole response}
\label{sec:four}

The images of a close binary and of a single source can be visually very
similar because both contain rings, multiple peaks, and diffraction
fringes generated by black-hole scattering. To isolate the information
that is intrinsic to the binary, we compare it with a single source that
has the same total brightness and is placed at the brightness centroid of
the pair. In this section we first define this reference image, then
expand the exact binary image about the centroid, and finally identify
the second central moment as the leading nontrivial source-structure
contribution. This construction also provides a quantitative residual
with which the numerical results can be tested.

\subsection{Single-source response and brightness-centroid reference}

For fixed black-hole parameters, observation direction, and lens aperture,
we retain the frequency dependence explicitly and denote by
\begin{equation}
 K(\alpha,\beta;x_{s};\omega)
 \label{eq:single_source_kernel}
\end{equation}
the unnormalized image-plane intensity produced by a point source of
unit brightness at transverse position $x_{s}$ and frequency $\omega$. The response function
$K$ contains the complete partial-wave propagation, black-hole
scattering, aperture diffraction, and Fourier imaging described in the
previous sections. Its first two arguments are image coordinates,
whereas $x_s$ is a source coordinate and $\omega$ is the angular frequency.
In the frequency scans below, the source positions and brightnesses are
held fixed, so that $x_c$ and $\mu_2$ do not vary with frequency.

The two sources are located at $x_{1}=a$ and $x_{2}=-a$, with
brightnesses $F_{1}$ and $F_{2}$, respectively. Since the sources are
mutually incoherent, Eq.~\eqref{eq:incoherent_image_sum} can be written
as
\begin{equation}
 I_{\mathrm{binary}}(\alpha,\beta;\omega)
 =F_{1}K(\alpha,\beta;a;\omega)
 +F_{2}K(\alpha,\beta;-a;\omega).
 \label{eq:binary_kernel_sum}
\end{equation}
We introduce the brightness ratio and total brightness
\begin{equation}
 q=\frac{F_{2}}{F_{1}}
 =\left|\frac{\mathcal{J}_{2}}{\mathcal{J}_{1}}\right|^{2},
 \qquad
 F_{\mathrm{tot}}=F_{1}+F_{2}.
 \label{eq:brightness_ratio}
\end{equation}
Thus $q$ is an intensity ratio rather than a field-amplitude ratio.

The natural position of the reference source is the brightness centroid
of the source pair,
\begin{equation}
 x_{c}=\frac{F_{1}x_{1}+F_{2}x_{2}}{F_{1}+F_{2}}
 =a\frac{1-q}{1+q}.
 \label{eq:brightness_centroid}
\end{equation}
We use a single source of brightness $F_{\mathrm{tot}}$ placed at
$x_c$ as the reference model,
\begin{equation}
 I_{\mathrm{ref}}(\alpha,\beta;\omega)
 =F_{\mathrm{tot}}K(\alpha,\beta;x_{c};\omega).
 \label{eq:reference_image}
\end{equation}
The binary source and this reference source have the same total
brightness and the same brightness centroid. Their difference therefore
does not measure a change in the total flux or an overall displacement,
but instead isolates the internal spatial structure of the source pair.
For an ideal single point source at $x_s$, choosing its own brightness
centroid, $x_c=x_s$, makes every central moment of order $n\geq1$ vanish,
\begin{equation*}
 \mu_n^{\mathrm{single}}=(x_s-x_c)^n=0.
\end{equation*}
This statement concerns moments of the \emph{source-brightness
distribution}, not moments or visible peaks of its image: a single source
may still generate rings and interference fringes after black-hole
scattering.

\subsection{Centroid expansion and second-order response}

We expand the single-source response with respect to the source position
about $x_s=x_c$ at each fixed frequency,
\begin{equation}
 K(\alpha,\beta;x_s;\omega)
 =K_c(\omega)+(x_s-x_c)K'_c(\omega)
 +\frac{1}{2}(x_s-x_c)^2K''_c(\omega)+\cdots,
 \label{eq:kernel_taylor_expansion}
\end{equation}
where
\begin{equation}
 \begin{aligned}
 K_c(\omega)&=K(\alpha,\beta;x_c;\omega),\\
 K'_c(\omega)&=\left.\frac{\partial K(\alpha,\beta;x_s;\omega)}{\partial x_s}\right|_{x_s=x_c},\\
 K''_c(\omega)&=\left.\frac{\partial^2K(\alpha,\beta;x_s;\omega)}{\partial x_s^2}\right|_{x_s=x_c}.
 \end{aligned}
 \label{eq:kernel_source_derivatives}
\end{equation}
These derivatives are taken with respect to the source position $x_s$,
with $\omega$, $\alpha$, and $\beta$ held fixed, not with respect to
the image coordinate $\alpha$. In $K_c(\omega)$ and its derivatives,
the image-coordinate arguments are suppressed for brevity.

Substitution into Eq.~\eqref{eq:binary_kernel_sum} gives
\begin{align}
 I_{\mathrm{binary}}(\alpha,\beta;\omega)
 ={}&F_{\mathrm{tot}}K_c(\omega)
 +\left[F_{1}(a-x_c)+F_{2}(-a-x_c)\right]K'_c(\omega)
 \nonumber\\
 &+\frac{1}{2}
 \left[F_{1}(a-x_c)^2+F_{2}(-a-x_c)^2\right]K''_c(\omega)
 +\cdots.
 \label{eq:binary_centroid_expansion}
\end{align}
By the definition of the brightness centroid,
\begin{equation}
 F_{1}(a-x_c)+F_{2}(-a-x_c)=0.
 \label{eq:first_moment_cancellation}
\end{equation}
The first-order term therefore vanishes exactly. It represents only a
translation of the source distribution and contains no intrinsic
binary-source structure once the reference source is placed at the
brightness centroid.

After the first-order translation has been removed, the next coefficient
is the second central moment of the source-brightness distribution,
defined by
\begin{equation}
 \mu_{2}=\frac{
 F_{1}(a-x_c)^2+F_{2}(-a-x_c)^2
 }{F_{\mathrm{tot}}}.
 \label{eq:second_central_moment_definition}
\end{equation}
Using Eqs.~\eqref{eq:brightness_ratio} and
\eqref{eq:brightness_centroid}, it becomes
\begin{equation}
\mu_{2}=\frac{4a^{2}q}{(1+q)^{2}}.
 \label{eq:second_central_moment}
\end{equation}
Consequently, the binary image takes the form
\begin{equation}
 I_{\mathrm{binary}}(\alpha,\beta;\omega)
 =F_{\mathrm{tot}}K(\alpha,\beta;x_c;\omega)
 +\frac{F_{\mathrm{tot}}\mu_{2}}{2}
 \left.\frac{\partial^{2}K(\alpha,\beta;x_s;\omega)}{\partial x_s^{2}}
 \right|_{x_s=x_c}
 +O(a^{3}).
 \label{eq:binary_second_order_expansion}
\end{equation}
The leading nontrivial difference from the reference image is therefore
\begin{equation}
 \Delta I_{\mathrm{binary}}(\alpha,\beta;\omega)
 \simeq
 \frac{2F_{\mathrm{tot}}a^{2}q}{(1+q)^{2}}
 \left.\frac{\partial^{2}K(\alpha,\beta;x_s;\omega)}{\partial x_s^{2}}
 \right|_{x_s=x_c}
 .
 \label{eq:second_order_image_response}
\end{equation}
We refer to Eq.~\eqref{eq:second_order_image_response} as the
second-order multipole response. It is the image-plane response to the
second central moment of the source-brightness distribution; it should
not be confused with a gravitational multipole moment of the
Schwarzschild spacetime.

\subsection{Structure measure, scaling laws, and range of validity}

To quantify the difference between the binary image and its optimal
single-source reference, we define
\begin{equation}
 S(\omega)=\frac{
 \left\|I_{\mathrm{binary}}(\omega)-I_{\mathrm{ref}}(\omega)\right\|_{2}
 }{
 \left\|I_{\mathrm{binary}}(\omega)\right\|_{2}
 },
 \label{eq:structure_strength}
\end{equation}
where the image-coordinate arguments are suppressed and the norm is
evaluated over the selected image region. In the
small-separation limit,
\begin{equation}
 S(\omega)\simeq
 \frac{\mu_{2}}{2}
 \frac{
 \left\|\left.\partial^{2}K(\alpha,\beta;x_s;\omega)/\partial x_s^{2}
 \right|_{x_s=x_c}\right\|_{2}
 }{
 \left\|K(\alpha,\beta;x_c;\omega)\right\|_{2}
 }.
 \label{eq:structure_strength_approximation}
\end{equation}
For a fixed imaging system, the ratio of the two kernel norms is fixed.
The leading parameter dependences are therefore
\begin{equation}
S\propto a^{2},
 \qquad
S\propto\frac{q}{(1+q)^{2}}.
 \label{eq:structure_scalings}
\end{equation}
The response is strongest for equal brightness, $q=1$, and tends to
zero for $q\rightarrow0$ or $q\rightarrow\infty$, where the image
reduces to that of the brighter source.

For an equal-brightness pair,
\begin{equation}
 q=1,
 \qquad x_c=0,
 \qquad \mu_2=a^2,
 \label{eq:equal_brightness_moment}
\end{equation}
and hence
\begin{equation}
 I_{\mathrm{binary}}(\alpha,\beta;\omega)
 =F_{\mathrm{tot}}K(\alpha,\beta;0;\omega)
 +\frac{F_{\mathrm{tot}}a^2}{2}
 \left.\frac{\partial^2K(\alpha,\beta;x_s;\omega)}{\partial x_s^2}\right|_{x_s=0}
 +O(a^4).
 \label{eq:equal_brightness_expansion}
\end{equation}
All odd central moments vanish in this symmetric case, so the next
correction is fourth order rather than third order.

The scaling relations above follow from a local expansion and therefore
have a definite domain of validity. The two sources must be sufficiently
close that
the single-source image response is smooth over the interval containing
them. A geometric small-separation condition is
\begin{equation}
 \frac{a}{z_0}\ll1.
 \label{eq:geometric_smallness}
\end{equation}
In addition, the displacement-induced phase variation across the lens
should remain small,
\begin{equation}
 \eta\equiv\frac{\omega d a}{z_0}\ll1.
 \label{eq:phase_smallness}
\end{equation}
Within this regime, the leading binary-source information relative to
the brightness-centroid single-source model is controlled by the second
central moment in Eq.~\eqref{eq:second_central_moment}.

\section{Numerical verification}
\label{sec:numerical_verification}

We now test the brightness-centroid expansion with the full partial-wave
solution followed by Fourier imaging.  Geometrized units with $M=1$ are
used throughout the numerical calculation.  The common geometrical
parameters are $z_0=6M$,
$r_{\mathrm{obs}}=20M$, $d=10M$, and $(\theta_0,\phi_0)=(0,0)$.  The
partial-wave sum is truncated at
$l_{\max}=\lceil 9(M\omega+1)\rceil$. The complex lens fields generated by
the two sources are imaged separately, and their image-plane intensities
are then added, as required for mutually incoherent sources.  Images used
only for visualization are normalized panel by panel and displayed with a
power-law contrast enhancement.  By contrast, all residuals, norms, and
structure measures reported below are evaluated from the original
image-plane intensities without an independent normalization of the two
images being compared.  We first examine the source-parameter dependence
at $M\omega=12$ ($l_{\max}=117$), and then vary the frequency at fixed
source parameters.

\subsection{Second-order response and parameter dependence}
\label{subsec:second_order_response}

\begin{figure}[tbp]
 \centering
 \includegraphics[width=0.42\textwidth]{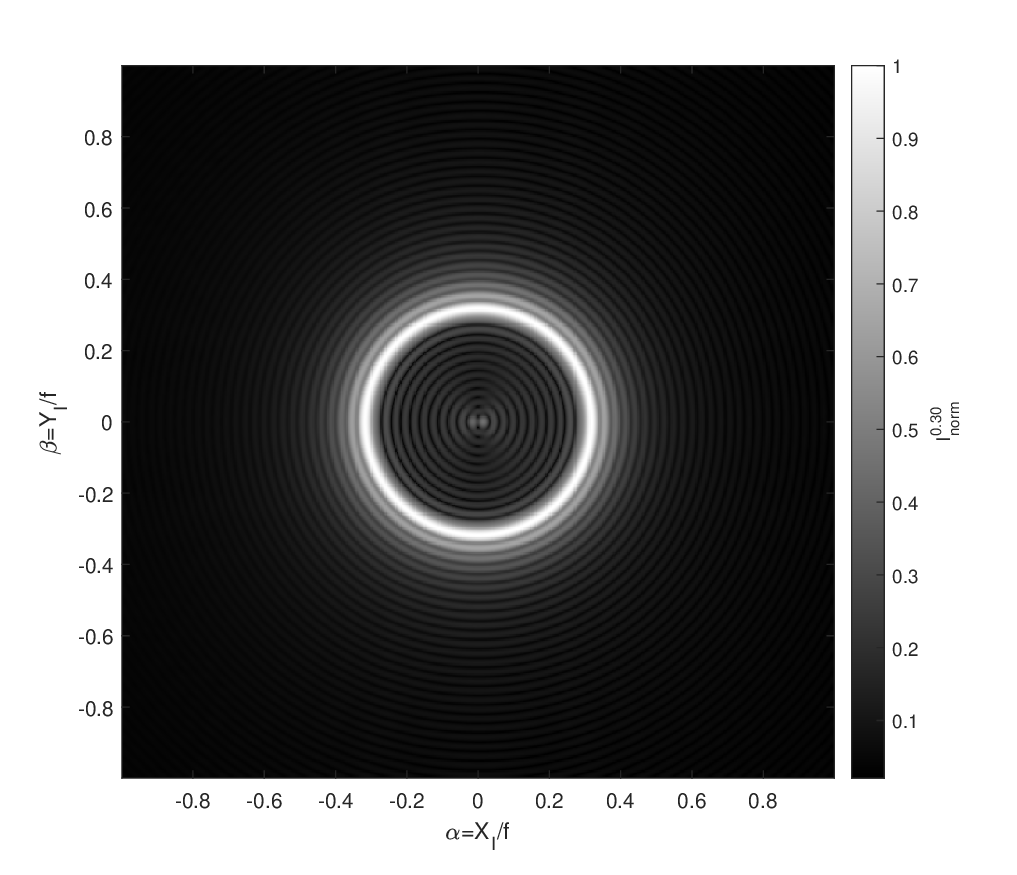}
 \includegraphics[width=0.42\textwidth]{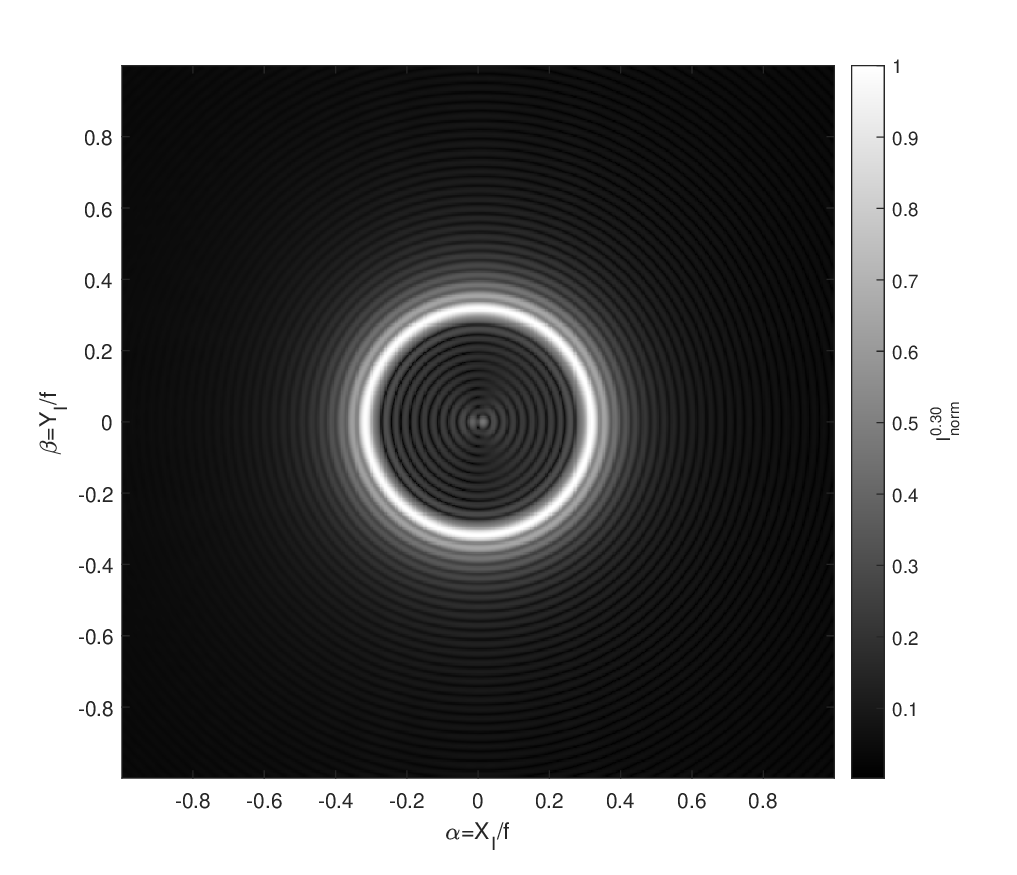}
 \caption{Normalized wave-optical images for two different source
 configurations.  Left: only the brighter source is present
 ($A_1=0$, $A_2=2$).  Right: the same bright source is accompanied by a
 weaker nearby source ($A_1=1$, $A_2=2$, hence $q=4$).  The common
 parameters are $M\omega=12$, $z_0=6M$, $a=0.1M$,
 $(\theta_0,\phi_0)=(0,0)$, $r_{\mathrm{obs}}=20M$, and $d=10M$.
 Each panel is normalized independently and displayed as
 $I_{\mathrm{norm}}^{0.30}$.  Their close
 visual similarity shows that the binary structure cannot in general be
 identified by counting bright rings or peaks alone.}
 \label{fig:single_binary_visual_comparison}
\end{figure}

Figure~\ref{fig:single_binary_visual_comparison} first illustrates why the
number of visible peaks is not a reliable indicator of the number of
sources.  The left panel contains only the brighter source,
$A_1=0$ and $A_2=2$, whereas the right panel includes a weaker nearby
companion, $A_1=1$ and $A_2=2$.  Since the source brightness is proportional
to $|A_s|^2$, the corresponding brightness ratio is $q=4$.  Although the
physical source configurations are different, the two normalized images
are visually very similar.  In particular, a single source already
produces a bright ring and a family of diffraction fringes.  Consequently,
ring multiplicity or one-dimensional peak counting can confuse
single-source diffraction structure with genuine binary-source structure.
This observation motivates the comparison with the equal-total-brightness
reference source located at the brightness centroid.

\begin{figure}[tbp]
 \centering
 \includegraphics[width=0.95\textwidth]{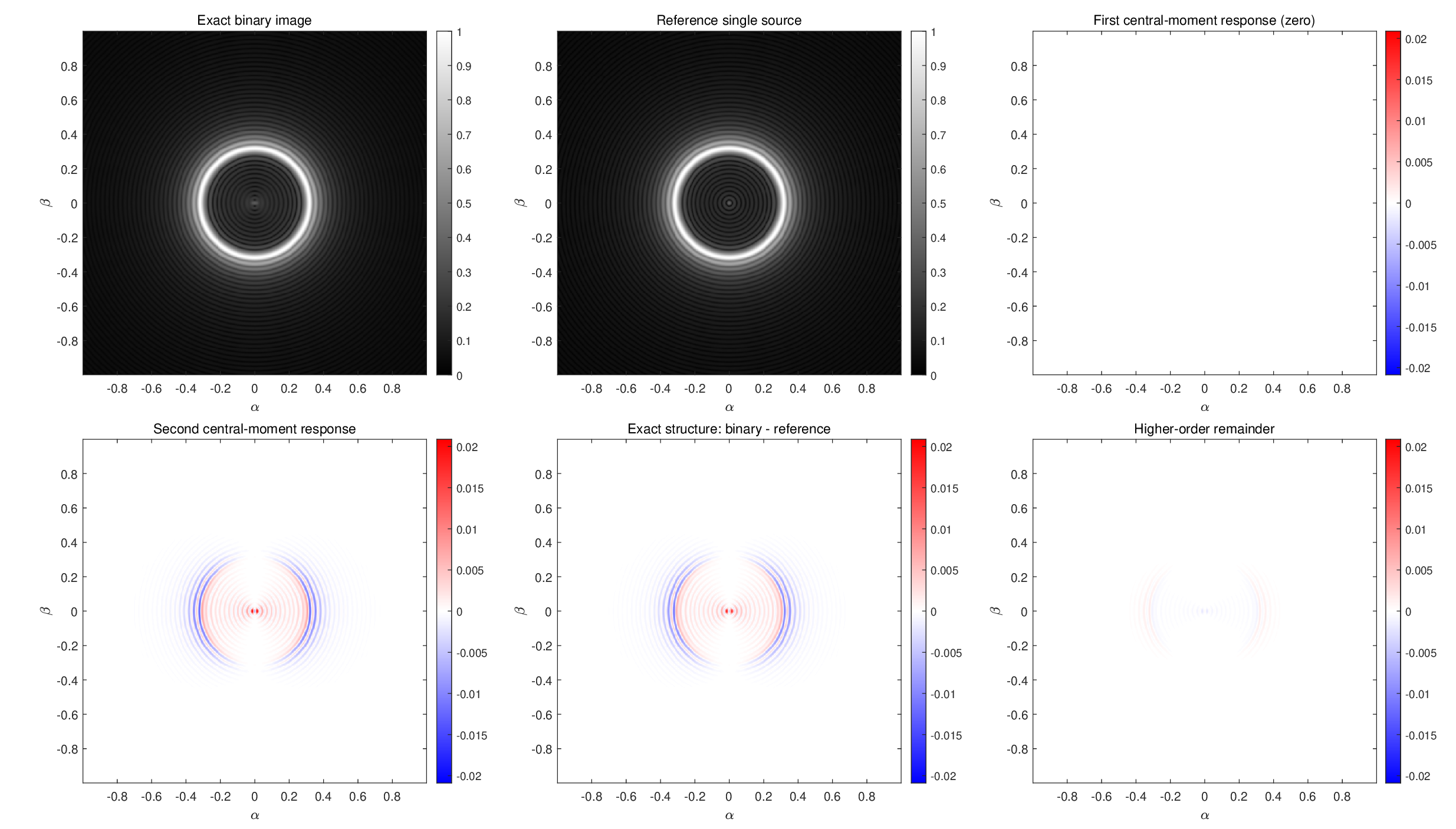}
 \caption{Central-moment decomposition for an equal-brightness binary.
 The panels show the exact binary image, the equal-total-brightness
 reference image at the brightness centroid, the first-order response, the
 second-order response, the exact binary-minus-reference residual, and the
 remaining higher-order contribution.  The source parameters are
 $a=0.05M$ and $A_1=A_2=1$ ($q=1$); the common imaging parameters are
 $M\omega=12$, $z_0=6M$, $(\theta_0,\phi_0)=(0,0)$,
 $r_{\mathrm{obs}}=20M$, and $d=10M$.  The first-order response vanishes,
 and the second-order response reproduces the dominant morphology of the
 exact residual.}
 \label{fig:equal_binary_decomposition}
\end{figure}

The central-moment decomposition is examined directly in
Fig.~\ref{fig:equal_binary_decomposition} for an equal-brightness binary.
The exact binary image and the brightness-centroid reference image are
nearly indistinguishable on their common intensity scale.  The first-order
response vanishes numerically, in agreement with
Eq.~\eqref{eq:first_moment_cancellation}.  The second-order response has the
same dominant morphology as the exact residual
$I_{\mathrm{binary}}-I_{\mathrm{ref}}$, while the remaining difference is
shown separately.  Morphological agreement alone does not quantify the
relative approximation error, which is examined in
Sec.~\ref{subsec:frequency_dependence}.  For this reflection-symmetric configuration all odd central
moments vanish; hence the leading omitted correction is fourth order in
$a$, rather than third order.

\begin{figure}[tbp]
 \centering
 \includegraphics[width=0.95\textwidth]{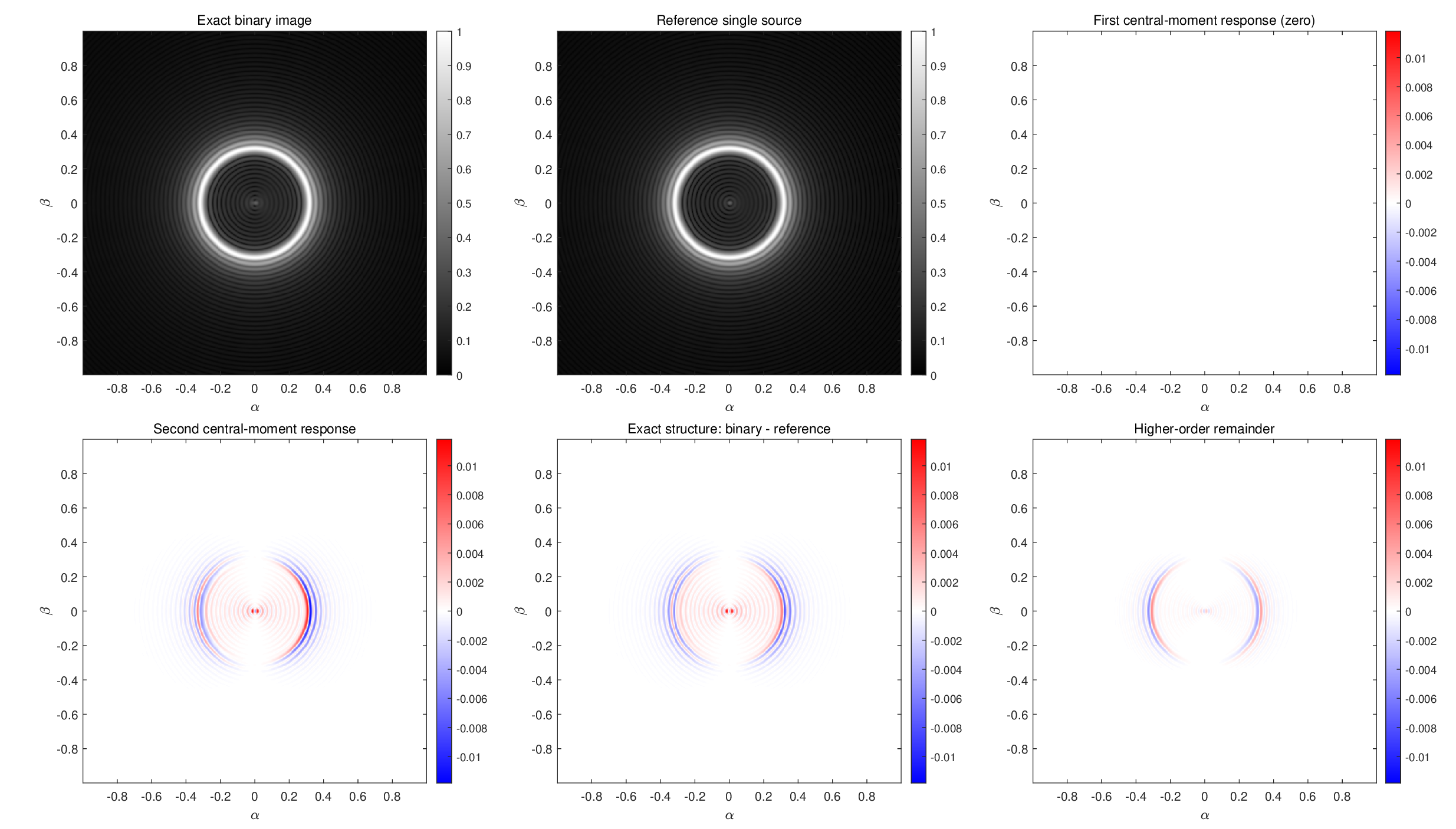}
 \caption{Central-moment decomposition for an unequal-brightness binary.
 The expansion is performed about the brightness centroid, so the
 first-order response still vanishes.  Here $a=0.05M$, $A_1=1$, and
 $A_2=2$ ($q=4$), while $M\omega=12$, $z_0=6M$,
 $(\theta_0,\phi_0)=(0,0)$, $r_{\mathrm{obs}}=20M$, and $d=10M$.
 The second-order response captures
 characteristic features of the binary-source structure; this comparison
 does not establish a prescribed quantitative accuracy.  The absence of reflection
 symmetry produces a larger higher-order remainder than in
 Fig.~\ref{fig:equal_binary_decomposition}.}
 \label{fig:unequal_binary_decomposition}
\end{figure}

Figure~\ref{fig:unequal_binary_decomposition} repeats the test for an
unequal-brightness pair.  The reference source is displaced from the
geometric midpoint to the brightness centroid $x_c$ given by
Eq.~\eqref{eq:brightness_centroid}.  With this choice the first central
moment again vanishes, even though the source distribution is no longer
reflection symmetric.  The second-order term continues to reproduce the
principal ring distortion in the exact residual.  The higher-order
remainder is more visible than in the equal-brightness case because the
third and higher odd central moments no longer vanish by symmetry.

\begin{figure}[tbp]
 \centering
 \includegraphics[width=0.5\textwidth]{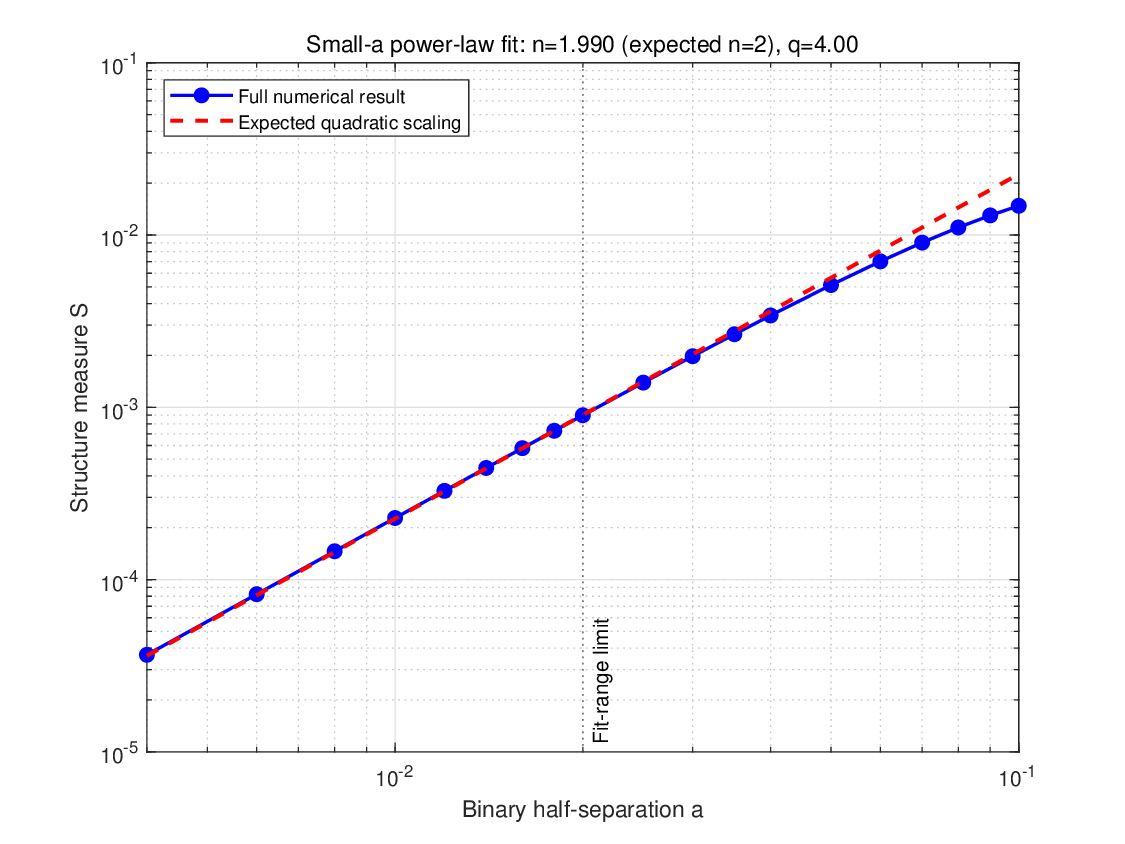}
 \caption{Structure measure $S$ as a function of the binary
 half-separation $a$ for $q=4$.  The small-separation fit gives
 $S\propto a^{1.990}$, consistent with the predicted quadratic scaling.
 The calculation uses $q=4$, $M\omega=12$, $z_0=6M$,
 $(\theta_0,\phi_0)=(0,0)$, $r_{\mathrm{obs}}=20M$, and $d=10M$;
 the scanned range is $0.004M\leq a\leq0.1M$.
 The fitted exponent is obtained only from the points to the left of the
 vertical dotted line; the remaining points display the onset of
 higher-order corrections.}
 \label{fig:a_scaling}
\end{figure}

We next test the separation dependence of the structure measure $S$
defined in Eq.~\eqref{eq:structure_strength}.  Figure~\ref{fig:a_scaling}
shows the full numerical result for $q=4$.  A linear fit to
$\log S$ as a function of $\log a$, restricted to the small-separation
points satisfying the phase-smallness criterion, gives
$S\propto a^{1.990}$.  This exponent is consistent with the quadratic
prediction in Eq.~\eqref{eq:structure_scalings}.  Only the points to the
left of the vertical dotted line are used in the fit.  At larger
separations, the numerical curve departs progressively from the quadratic
law because higher central moments and higher-order lens-phase variations
are no longer negligible.

\begin{figure}[tbp]
 \centering
 \includegraphics[width=0.5\textwidth]{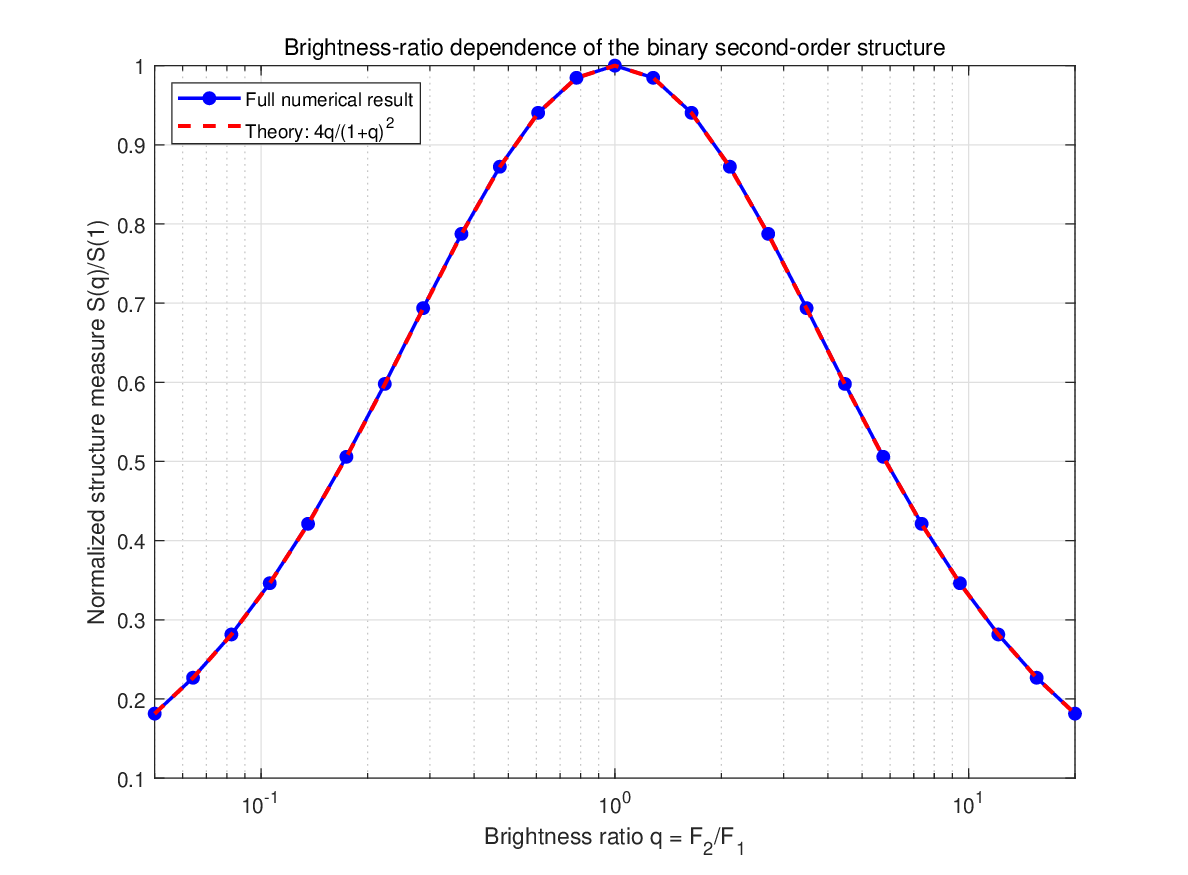}
 \caption{Normalized structure measure as a function of the brightness
 ratio $q=F_2/F_1$.  The numerical result is compared with
 $S(q)/S(1)=4q/(1+q)^2$.  The response is maximal for equal brightness,
 decreases when either source dominates, and is invariant under
 $q\leftrightarrow 1/q$.  The fixed parameters are $M\omega=12$,
 $z_0=6M$, $a=0.05M$, $(\theta_0,\phi_0)=(0,0)$,
 $r_{\mathrm{obs}}=20M$, and $d=10M$, and the brightness ratio is scanned
 over $0.05\leq q\leq20$.}
 \label{fig:q_scaling}
\end{figure}

Finally, Fig.~\ref{fig:q_scaling} tests the brightness-ratio dependence at
fixed separation.  Both the numerical structure measure and the
theoretical curve are divided by their values at $q=1$.  The prediction
following from Eq.~\eqref{eq:second_central_moment}, when the second-order
approximation is adequate and the kernel norm ratio varies weakly with
the centroid position, is
\begin{equation}
 \frac{S(q)}{S(1)}
 \simeq
 \frac{4q}{(1+q)^2}.
 \label{eq:normalized_q_scaling}
\end{equation}
The numerical points closely follow this parameter-free shape.  The
response is maximal at equal brightness and decreases when either source
dominates.  The symmetry under $q\leftrightarrow 1/q$ is also reproduced.
Here the value at $q=1$ fixes the common vertical normalization; thus the
comparison tests the predicted dependence on $q$, rather than fitting an
additional amplitude at every data point.  Agreement of this scalar
measure with the predicted shape does not by itself imply that the full
residual image is accurately reproduced by the second-order response.

\FloatBarrier

\subsection{Frequency dependence and validity of the second-order approximation}
\label{subsec:frequency_dependence}

We now fix the binary half-separation at $a=0.05M$ and the brightness
ratio at $q=4$ and investigate the frequency dependence of the
second-order response, the full residual strength, and the accuracy of
the second-order approximation. The remaining parameters are unchanged.

\begin{figure}[tbp]
 \centering
 \includegraphics[width=0.98\textwidth]{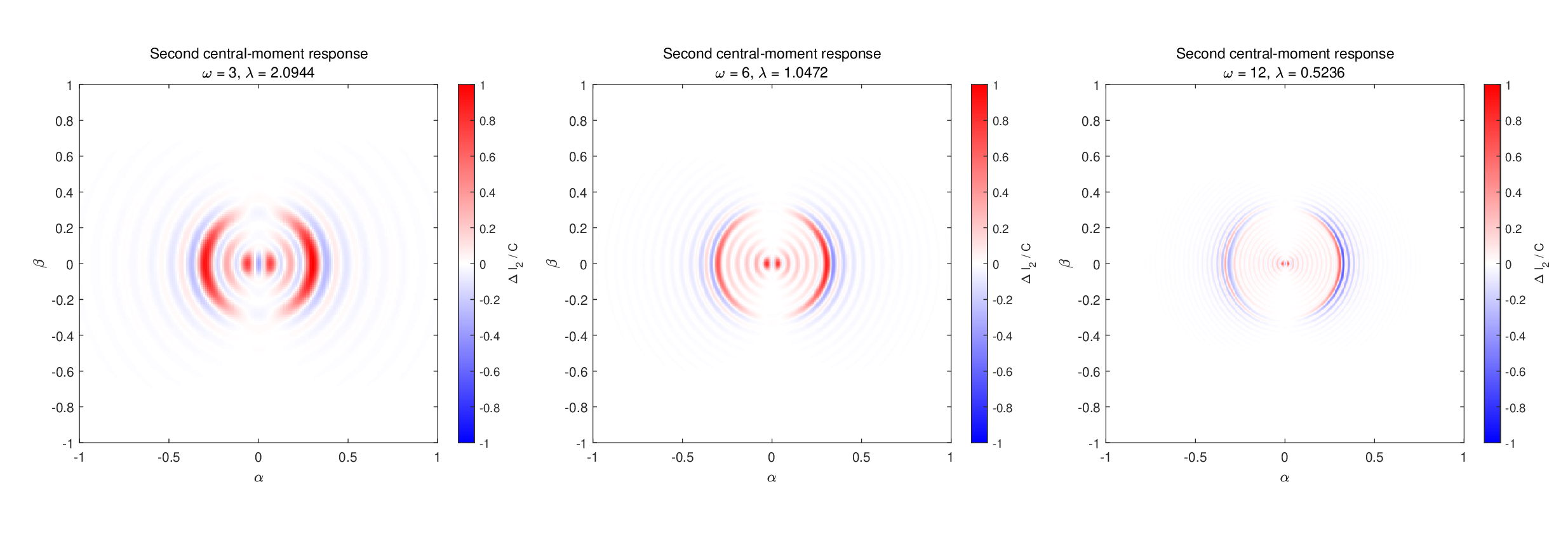}
 \caption{Second-order central-moment response at different frequencies.
 From left to right, $M\omega=3,6,12$, corresponding to
 $\lambda/M=2.0944,1.0472,0.5236$. The source parameters are fixed at
 $a=0.05M$ and $q=4$, with the remaining parameters as specified earlier
 in this section. Each panel shows $\Delta I_2/C$, where $C$ is the
 maximum of $|\Delta I_2|$ over the displayed regions at all three
 frequencies.}
 \label{fig:frequency_second_order}
\end{figure}

Figure~\ref{fig:frequency_second_order} shows the second-order response at
$M\omega=3,6,12$. As the frequency increases, the spatial fringes become
finer and the response morphology changes. The second central moment of
the source remains fixed in this calculation, so these changes arise
from the frequency dependence of the imaging kernel. The three panels
use a common display scale and show the second-order response itself;
its accuracy in describing the full image residual is examined below.

\begin{figure}[tbp]
 \centering
 \includegraphics[width=0.95\textwidth]{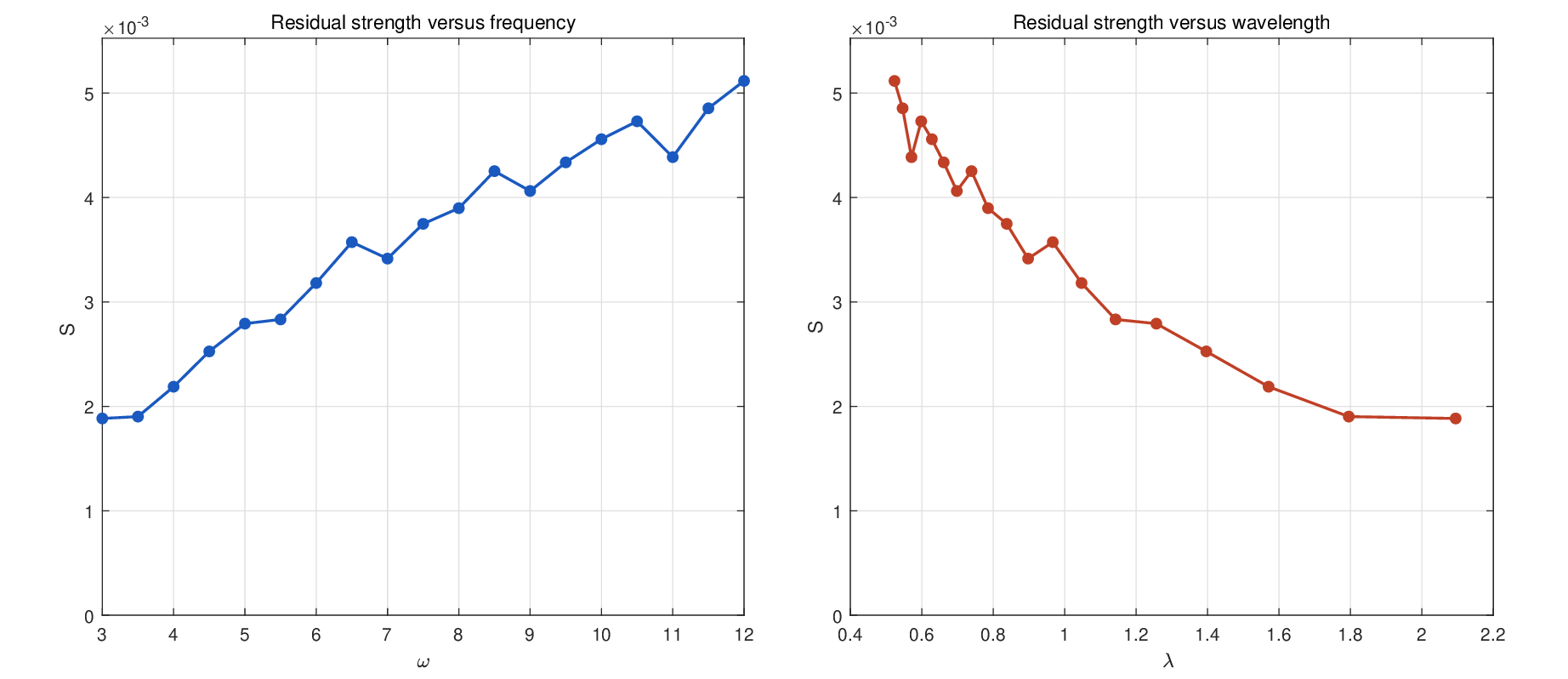}
 \caption{Full residual strength $S$ versus frequency (left) and
 wavelength (right), with $a=0.05M$ and $q=4$. The frequency scan covers
 $3\leq M\omega\leq12$ in steps of $0.5$. Both panels show the same
 data, related by $\lambda/M=2\pi/(M\omega)$; the plotted axes use
 $M=1$.}
 \label{fig:frequency_strength}
\end{figure}

Figure~\ref{fig:frequency_strength} shows the full residual strength $S$
as a function of frequency and wavelength. Over the scan
$3\leq M\omega\leq12$, $S$ increases overall from approximately
$1.9\times10^{-3}$ to $5.1\times10^{-3}$, with local variations.
Thus, for the parameters considered, the relative difference between the
binary image and the reference single-source image generally becomes
stronger as the frequency increases and the wavelength decreases.
Here $S$ is calculated from the full image residual and includes both
second-order and higher-order contributions. Its increase therefore does
not imply an improvement in the accuracy of the second-order approximation.

To quantify how accurately the second-order response describes the full
image residual, we define
\begin{equation}
 E_2(\omega)=
 \frac{\|\Delta I-\Delta I_2\|_2}{\|\Delta I\|_2},
 \qquad \Delta I=I_{\mathrm{binary}}-I_{\mathrm{ref}}.
 \label{eq:second_order_relative_error}
\end{equation}
The numerator measures the higher-order remainder omitted by the
second-order truncation, while the denominator measures the full image
residual. When $E_2\ll1$, the second-order response accurately describes
the full residual. When $E_2$ reaches order unity, the higher-order
remainder is comparable in norm to the full residual, and the
second-order truncation no longer provides a description with a small
relative error.

\begin{figure}[tbp]
 \centering
 \includegraphics[width=0.95\textwidth]{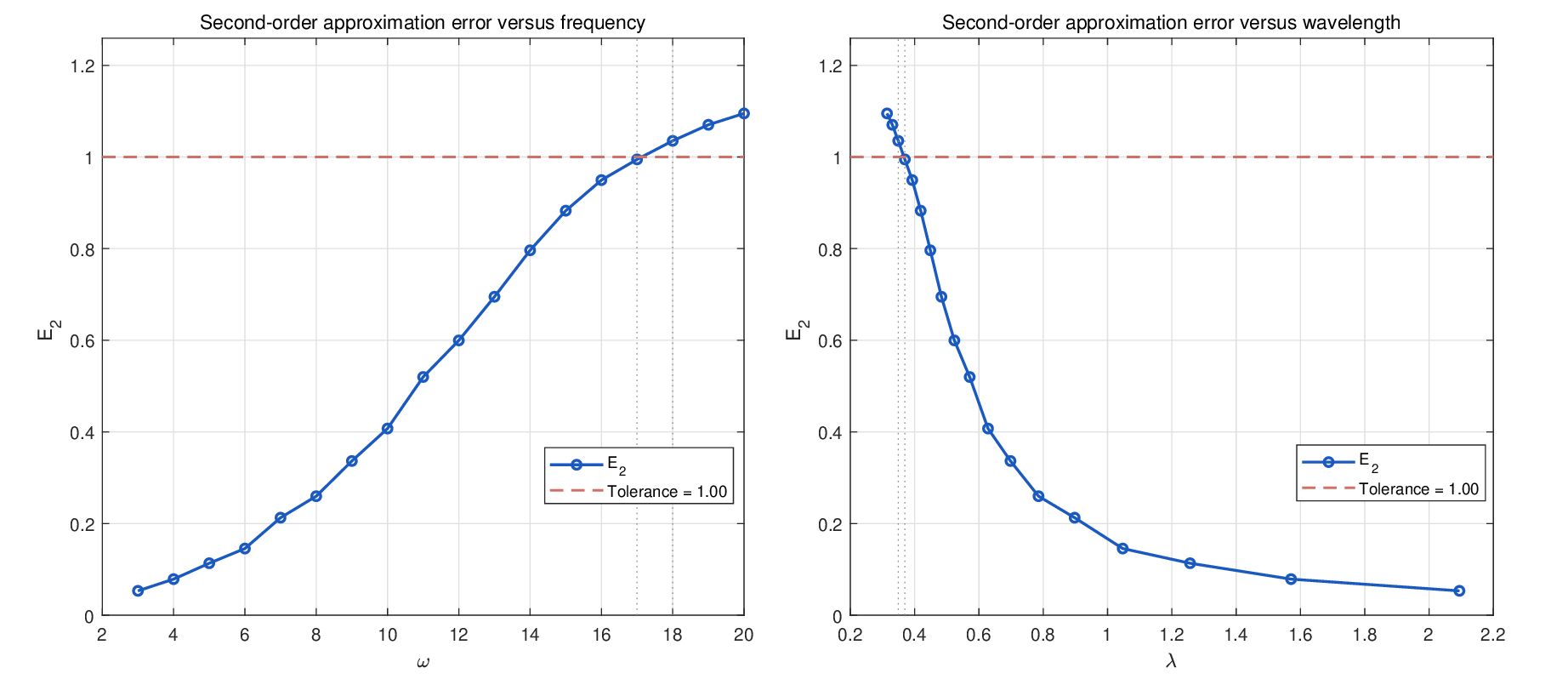}
 \caption{Relative second-order approximation error $E_2$ versus
 frequency (left) and wavelength (right), with $a=0.05M$ and $q=4$.
 The frequency scan covers $3\leq M\omega\leq20$ in steps of $1$.
 The horizontal dashed line marks $E_2=1$, and the vertical dotted lines
 bracket the first upward crossing, $17<M\omega<18$, and its
 corresponding wavelength interval. The plotted axes use $M=1$.}
 \label{fig:frequency_error}
\end{figure}

Figure~\ref{fig:frequency_error} shows the relative error over
$3\leq M\omega\leq20$. As the frequency increases, $E_2$ rises from
approximately $0.053$ to $1.095$. At the three frequencies selected in
Fig.~\ref{fig:frequency_second_order}, the corresponding errors are
approximately $0.053$, $0.145$, and $0.600$, respectively. Thus, although
the second-order response can be calculated and displayed at each
frequency, its accuracy as an approximation to the full residual
decreases with increasing frequency in this scan.

Using $E_2=1$ to mark where the relative error reaches unity, we find
$E_2(17)\simeq0.995$ and $E_2(18)\simeq1.035$. The first upward crossing
is therefore bracketed by
\[
 17<M\omega<18,
\]
with the corresponding wavelength interval
\[
 0.3491<\lambda/M<0.3696.
\]
This result identifies the frequency range in which the second-order
approximation error reaches unity for the specified source parameters
and imaging configuration. It indicates that higher-order contributions
cannot be neglected, rather than that the second-order response itself
vanishes. Conversely, $E_2<1$ does not by itself imply a high-accuracy
approximation.

\FloatBarrier

\section{Conclusions}
\label{sec:conclusions}

We have studied the wave-optical imaging of two nearby, mutually
incoherent point sources by a Schwarzschild black hole. Using the
spherical-harmonic addition theorem, we constructed the partial-wave
fields for arbitrary source directions and added their intensities after
performing Fourier imaging separately. By introducing a reference single
source with the same total source brightness at the brightness centroid,
we analyzed the image differences arising from the internal spatial
distribution of the binary source.

In an expansion about the brightness centroid, the first-central-moment
contribution vanishes exactly, while the second central moment provides
the leading structural correction to the image. Within the applicable
regime, the numerical results support the quadratic dependence of the
residual strength on half-separation and the brightness-ratio dependence
predicted by the second central moment. The second-order response
accounts for characteristic features of the residual structure, while
its quantitative accuracy must be assessed by comparing the full
residual with the second-order prediction. For fixed source parameters,
the central moments of the brightness distribution are frequency
independent; the wavelength dependence of the image differences
therefore resides entirely in the imaging kernel, which encodes
propagation, black-hole scattering, aperture diffraction, and Fourier
imaging.

The wavelength of the incident wave controls how distinguishable a
nearby pair is from a single source. The relevant control parameter is
the displacement-induced phase accumulated across the aperture,
$\eta=\omega da/z_0$. In the long-wavelength regime, $\eta\ll1$, the
imaging kernel varies little over the source separation, the
central-moment corrections are strongly suppressed, and the binary image
reduces to that of a single source placed at the brightness centroid:
the internal structure is effectively invisible. As the wavelength
decreases, the fringes of the kernel become finer, and a fixed source
displacement produces a progressively larger fractional change of the
image. Over the scan $3\leq M\omega\leq12$, corresponding to
$2.09\geq\lambda/M\geq0.52$, the full residual strength grows from
approximately $1.9\times10^{-3}$ to $5.1\times10^{-3}$, with local
variations. Throughout this scan the angular separation of the pair,
$2a/z_0\simeq1.7\times10^{-2}$, remains below the diffraction scale
$\sim\lambda/d$ of the finite aperture, so peak counting and
Rayleigh-type criteria would classify the pair as an unresolved single
source. The growing residual shows that binary information is
nevertheless retained and increasingly emphasized at shorter
wavelengths; accessing it requires the model-based residual rather than
visual separation of two peaks.

Distinguishability and interpretability do not improve in step. Over
$3\leq M\omega\leq20$, the relative error of the second-order
approximation rises from approximately $0.053$ to $1.095$, and its first
upward crossing of unity is bracketed by $17<M\omega<18$, or
$0.3491<\lambda/M<0.3696$, for $a=0.05M$, $q=4$, and the imaging
configuration adopted here. In this short-wavelength regime the
displacement-induced phase varies appreciably across the aperture and
higher central moments of the source-brightness distribution contribute,
so the second-order response no longer provides a small relative error
and higher multipoles are required. Together, the two trends delineate
an intermediate wavelength regime in which the binary signature is
strong enough to be of practical interest while the second-order
description retains controlled accuracy; within this regime, shorter
wavelengths favor detectability, whereas approaching its short-wavelength
edge the second-order interpretation deteriorates.

These results provide a second-order description of the image differences
produced by nearby binary sources and quantify the wavelength dependence
of their distinguishability and of the range of validity of that
description. The present analysis is restricted to an ideal, noise-free
forward imaging model. Determining the number of sources from an unknown
observed image requires further consideration of parameter fitting,
detector response, and noise. Since the binary signature is chromatic,
suppressed at long wavelengths and increasingly structured at short
wavelengths, combining observations at different wavenumbers may improve
the identification of nearby sources beyond single-frequency limits.
Finite bandwidth, partially coherent sources, and rotating black-hole
backgrounds provide additional directions for extending this work.

\section*{Acknowledgments}
We would like to thank Dr. Yang Huang for numerous helpful discussions.
This work is supported by the National Natural Science
Foundation of China (NSFC) under Grant nos. 12235019, 12275106.

\end{document}